\documentclass[10pt,letterpaper]{article}
\usepackage[top=0.85in,left=2.75in,footskip=0.75in]{geometry}

\usepackage{changepage}
\usepackage{comment}

\usepackage[utf8]{inputenc}

\usepackage{textcomp,marvosym}

\usepackage{fixltx2e}

\usepackage{amsmath,amssymb}

\usepackage{cite}

\makeatletter
\renewcommand{\@biblabel}[1]{\quad#1.}
\makeatother

\usepackage{lastpage,fancyhdr,graphicx}
\usepackage{epstopdf}

\title{}
\author{}

\usepackage{color}

\newcommand{\BE}{\begin{equation}}
\newcommand{\EE}{\end{equation}}
\newcommand{\BA}{\begin{eqnarray}}
\newcommand{\EA}{\begin{eqnarray}}

\newcommand{\vx}{\vec x}
\newcommand{\vy}{\vec y}
\newcommand{\vv}{\vec v}

\begin{document}
\vspace*{0.35in}

\begin{flushleft}
{\Large \textbf\newline{Correlation networks from flows. The
case of forced and time-dependent advection-diffusion dynamics
}
}
\newline
\\

Liubov Tupikina\textsuperscript{1,2,*}, Nora
Molkenthin\textsuperscript{3}, Crist\'{o}bal
L\'{o}pez\textsuperscript{4}, Emilio
Hern\'{a}ndez-Garc\'{\i}a\textsuperscript{4}, Norbert
Marwan\textsuperscript{1},  J\"{u}rgen
Kurths\textsuperscript{1,2}

\bigskip
\bf{1} Potsdam Institute for Climate Impact Research, P.O. Box
601203, 14412 Potsdam, Germany
\\
\bf{2} Humboldt Universit{\"a}t zu Berlin, 10099  Berlin, Germany
\\
\bf{3} Department of Physics, Technical University of Darmstadt, 64289 Darmstadt, Germany 
\\
\bf{4} IFISC (CSIC-UIB), Instituto de F\'{\i}sica
Interdisciplinar y Sistemas Complejos, Campus Universitat de
les Illes Balears, E-07122, Palma de Mallorca, Spain.

\bigskip
* tupikina@pik-potsdam.de

\end{flushleft}

\section*{Abstract}
Complex network theory provides an elegant and powerful
framework to statistically investigate different types of
systems such as society, brain or the structure of local and
long-range dynamical interrelationships in the climate system.
Network links in climate networks typically imply information,
mass or energy exchange. However, the specific connection
between oceanic or atmospheric flows and the climate network's
structure is still unclear. We propose a theoretical approach
for verifying relations between the correlation matrix and the
climate network measures, generalizing previous studies and
overcoming the restriction to stationary flows. Our methods are
developed for correlations of a scalar quantity (temperature,
for example) which satisfies an advection-diffusion dynamics in
the presence of forcing and dissipation. 
Our approach reveals that  
correlation networks are not sensitive 
to steady sources and sinks 
and the profound impact of the signal decay rate on the network topology. 
We illustrate our
results with calculations of degree and clustering for a
meandering flow resembling a geophysical ocean jet.


\section*{Introduction}
\label{sec:intro}

The network approach has become an essential tool in the study
of complex systems
\cite{Boccaletti2006,Ghil2008,mugnolo2008}, 
where networks are reconstructed
from time series in order to uncover underlying dynamics 
\cite{Holme2012,Zamora2011,Timme2014, marwan2015, Gao2015}.
Climate networks, i.e. those in which geographical nodes are
linked when there is similar climatic dynamics on them (as
measured by correlations, mutual information, etc.), have been
thoroughly investigated in the last years in
\cite{Yamasaki2008, palus2011, Malik2011,Mheen2013,Wang2013, Deza2015}. 
In the same context of geophysical systems,  
flow networks have
also been introduced \cite{rossi2014, sergiacomi2015, Ser-giacomi2015, Ser-Giacomi2015pre}. 
They are networks in which geographical nodes are linked when
there is fluid transport from one location to another. Since
correlations between different regions of a flow or geophysical
system should be greatly influenced by the mass transport among
them, it is natural to search for the relationship between
these two types of networks, which will also help to understand
the meaning of some of the teleconnections appearing in the
climate network analysis. 
The works \cite{Molkenthin2014a} and
\cite{Rehfeld2014} are in this line, where networks were
constructed from flow systems using a continuous analogue of
the Pearson correlation.  However these approaches have
their limitations, mainly the restriction on the velocity
fields to be constant in time. But the time-dependency plays an
important role in real-world flows, for instance, all ocean
currents vary over a large range of time scales
\cite{Watts1982, Siedler2001, deser2004}. 

In this paper we investigate general relationships between
climate networks (specifically, networks built from
correlations) and flow networks. In particular we develop a
method for the analysis of time-dependent flows and demonstrate
its potential for a specific model describing a meandering
current. The quantity for which we compute spatial correlations
is a scalar which is transported by the flow following an
advection-diffusion dynamics. We can think on it as the
`temperature' of water in an ocean flow, but the formalism
would apply to any transported quantity that could be
considered `passive' in some range of time scales. To avoid
trivial homogenization, the scalar is forced by sources and
sinks, which have both a spatially-dependent constant component
and a time-varying stochastic part, and a decay process that
prevents indefinite build-up, finally dissipating the input
from the sources. By discretizing the system dynamics in space
and time we obtain a linear recursive equation for the
time-series of the scalar. We estimate the spatial correlation
matrix from the time-series by averaging over various realizations of
the noise. The correlation matrix can be thresholded, and
interpreted as the adjacency matrix of the correlation network,
which can then be analyzed using network measures which
provides understanding of the formal relationship between the
Lagrangian transport in the basic flows and the corresponding
correlation network as used in climate networks.

The paper is organized as follows: First in the section Methods 
we introduce the tools for 
the construction of networks from general time-dependent flows, and describe our
example meandering-jet model. 
The Results section describes the properties of our main formulae and 
illustrate them with the model flow. In the last section we
discuss the main findings of the paper.

\section*{Methods}
\label{sec:methods}

We introduce an algorithm for the construction of correlation
networks from the spatial distribution of a scalar (e.g.
`temperature') transported in a two-dimensional domain by an
advection-diffusion equation (ADE) with additional forcing and
decay terms:
\BE
\frac{\partial T(\vx,t)}{\partial t} = \kappa \Delta T(\vx,t) -
\vv(\vx,t)\cdot \nabla T(\vx,t) + F(\vx) - bT(\vx,t) +
\sqrt{D}\xi (\vx,t), \label{ademodtime}
\EE
where $\kappa$ is the diffusion coefficient, $\vv(\vx, t)$ is
the time-dependent bidimensional velocity field which we assume
to be incompressible, $F(\vx)$ is the forcing, which describes
time-independent sources and sinks, $\xi (\vx,t)$ is
uncorrelated Gaussian white noise with zero mean and
correlations $\langle
\xi(\vx,t)\xi(\vy,t')\rangle=\delta(t-t')\delta(\vx-\vy)$. $D$
is noise intensity and $b$ is a damping parameter which sets
the time-scale at which perturbations are dissipated in the
system. We add decay
and forcing to avoid convergence of the scalar distribution 
to a simple homogeneous equilibrium, and these processes are
actually present in real geophysical flows 
\cite{Hernandez-Garcia2010}.

\subsection*{Discretised dynamics}

The algorithm of network construction for a time-dependent
velocity field requires first a discretisation of
(\ref{ademodtime}). Let us consider first the simplified
equation without forcing and decay:
\begin{equation}
\frac{\partial T}{\partial t} = \kappa \Delta T - \vv(\vx,t)\cdot \nabla T.
\label{ademodtime1}
\end{equation}

We discretize (\ref{ademodtime1}) using an Euler scheme for a
regular $N\times N$-lattice with spatial resolution 
$\Delta x$ and time-interval $\Delta t$. The horizontal and
vertical components of velocity field for the lattice point
$(i,j)$ at time step $k=t/\Delta t$ are $v^x_{ij}(k)$ and
$v^y_{ij}(k)$. This gives:
\begin{eqnarray}
&& T_{ij} (k+1) = T_{ij} (k) - \nonumber \\
&& \frac{\Delta t}{2\Delta x} (v^x_{ij}(k) T_{i+1 j}(k)
-v^x_{ij}(k) T_{i-1 j}(k) +
v^y_{ij}(k) T_{i j+1}(k) - v^y_{ij}(k) T_{i j-1}(k))
+  \nonumber \\
&&\frac{\kappa \Delta t}{\Delta x^2} (T_{i j+1}(k)
+ T_{i j-1}(k) + T_{i+1 j}(k) + T_{i-1 j}(k) - 4T_{i j}(k)),
\label{dis}
\end{eqnarray}
where the node's indices are $i,j \in [1,N]$. We use open
boundary conditions. The discretisation parameters  $\Delta x$ and $\Delta t$ should fulfill 
the Courant-Friedrichs-Lewy condition \cite{press1988} 
for the stability of the discretisation scheme
$$
\frac{\kappa\Delta t}{\Delta x^2} << 1, \ \ \  \frac{{\rm max} (v(x,t)) \Delta t}{\Delta x} << 1 . 
$$
Equation (\ref{dis}) can be written in a matrix form in terms
of the $N^2\times N^2$ one-step transformation matrix
$\mathbf{P}(k)=\mathbf{P}(v_{ij}(k))$ for time step $k$ and the
$N^2$-dimensional state-vector $T(k)$ of components
$\left(T(k)\right)_{\vx}=T_{ij}(k)$, with $(i,j)$ the lattice
coordinates of $\vx$:
\begin{equation}
T(k+1) = \mathbf{P}(k)T(k).
\label{timestep}
\end{equation}
Iterating (\ref{timestep})
leads, for $k \ge k'$, to
\begin{equation}
T(k+1) = \mathbf{M}_{kk'}T(k'),
\label{iterat}
\end{equation}
where
\begin{equation}
\mathbf{M}_{kk'} = \mathbf{P}(k) \mathbf{P}(k-1) ... \mathbf{P}(k'+1)\mathbf{P}(k')
\label{timestep2}
\end{equation}
is the analogous to the transport matrix defining the flow
networks in \cite{sergiacomi2015}. 
Here it is computed from a discretization of the ADE, whereas
in other works \cite{Ser-giacomi2015, Ser-Giacomi2015pre} 
it is computed by the Ulam method that involves
the Lagrangian trajectories of particles, but the meaning is
the same: it is the matrix that evolves in time the vector
$T(k)$.

Adding the decay term $-bT$ to equation (\ref{ademodtime1}):
\begin{equation}
\frac{\partial T}{\partial t} = \kappa \Delta T - \vv(\vx,t)\cdot\nabla T -bT
\label{ademodtime2}
\end{equation}
does not pose technical difficulties, since the change of the
variables $T(k) = e^{-b\Delta t k} \widetilde{T}(k)$ reduces
Eq. (\ref{ademodtime2}) to (\ref{ademodtime1}) for
$\widetilde{T}(k)$. Therefore the one-step solution
(\ref{timestep}) becomes:
\begin{equation}
T(k+1) = e^{-b\Delta t}\mathbf{P}(k)T(k).
\label{timestep2}
\end{equation}
Being a transport matrix, the eigenvalue with largest modulus
of matrix $\mathbf P(k)$ is $1$. The new one-step
transformation $e^{-b\Delta t}\mathbf P(k)$ will have
eigenvalues which in modulus are smaller than 1, ensuring that
perturbations become damped.

Reintroducing the forcing terms $F(\vx)+ \sqrt{D}\xi(\vx,t)$ from
(\ref{ademodtime}) into the discretized framework (\ref{dis})
can be done for example by integrating them with the Euler
method. The one-step solution becomes
\begin{eqnarray}
T(k+1) = e^{-b\Delta t} \mathbf{P}(k)T(k) 
+ \Delta t F + s\epsilon (k).
\label{timestep1}
\end{eqnarray}
$F$ is the time independent spatial forcing vector, and
$\epsilon(k)$ is, at each time $k$, a vector of independent
Gaussian random variables of zero mean and unit variance. These
vectors are uncorrelated at different times. From the
stochastic Euler method \cite{toral2014}, the intensity of the
discretized noise is $s=\sqrt{D\Delta t/\Delta x^2}$. Iteration
of Eq. (\ref{timestep1}) for $(k+1)$ time steps gives the time
evolution of the scalar distribution vector:
\begin{equation}
T(k+1) = \mathbf{G}_{k0}~T(0) + \Delta t \sum_{l=0}^{k}\mathbf{G}_{kk+1-l} ~F + s \sum_{l=0}^{k}\mathbf{G}_{kk+1-l}~\epsilon(k-l)\ .
\label{inttimeser}
\end{equation}
We have introduced the propagation matrix, or propagator:
\begin{equation}
\mathbf{G}_{kk'} \equiv  e^{-b\Delta t} \mathbf{P}(k) e^{-b\Delta t} \mathbf{P}(k-1) ... e^{-b\Delta t} \mathbf{P}(k') =
e^{-(k+1-k')b\Delta t} \mathbf{M}_{kk'} \ , \ k \ge k' ,
\label{Gmatrix1}
\end{equation}
and for notational convenience, we have defined
\begin{equation}
\mathbf{G}_{k k+1} \equiv {\cal I} \ ,
\label{iden}
\end{equation}
the $N^2\times N^2$ identity matrix.

\subsection*{Calculation of correlations}

We are now able to compute the correlations associated to the
time series generated by Eq. ({\ref{inttimeser}). We consider
the direct product matrix $T(k)T(k)^\dagger$ (the superindex
$\dagger$ means transpose) whose matrix elements are products
of the transported field at different spatial points
$\left(T(k)T(k)^\dagger\right)_{\vx \vy}=T(k)_{\vx}T(k)^\dagger_{\vy}$.
We average it over realizations of the noise $\epsilon$,
operation which is denoted by $\langle .\rangle$. We also
include in the same operation averaging over the initial
condition $T(0)$, for which we assume $\langle T(0)\rangle=0$.
But we will see that in fact this assumption is irrelevant for
our results, since the final expressions at long times lose
dependence on the initial condition. Using
$\langle\epsilon(k)\epsilon(k')\rangle={\cal I}\delta_{kk'}$,
we find:
\begin{eqnarray}
&&\langle T(k+1)T(k+1)^\dagger \rangle =\mathbf{G}_{k0} \langle T(0)T(0)^\dagger \rangle \mathbf{G}_{k0}^\dagger + \nonumber \\
&& \left(\Delta t\right)^2\sum_{l=0}^k\sum_{l'=0}^k \mathbf{G}_{kk+1-l}FF^\dagger\mathbf{G}_{kk+1-l'}^\dagger +
s^2\sum_{l=0}^k \mathbf{G}_{kk+1-l}\mathbf{G}_{kk+1-l}^\dagger \ .
\label{correl1}
\end{eqnarray}
The first term in the r.h.s. of (\ref{correl1}) gives the
evolution of the initial correlations. Because of the
properties of the eigenvalues of $\mathbf{G}_{k0}$, this term
will decrease with $k$ and become negligible after a number $k$
of steps such that the corresponding time $k\Delta t$ satisfies
$bk\Delta t >>1$. In the same limit, by averaging Eq.
(\ref{inttimeser}), we see that
\begin{equation}
\langle T(k+1)\rangle = \Delta t\sum_{l=0}^k \mathbf{G}_{kk+1-l} F \ ,\ b\Delta t k >> 1\ ,
\end{equation}
so that the second term in the r.h.s. of Eq. (\ref{correl1}) is
$\langle T(k+1)\rangle \langle T(k+1)^\dagger\rangle$.
Combining these facts, we obtain for the spatial covariance of
the transported scalar, if $b k \Delta t >> 1$:
\begin{eqnarray}
\mathrm{Cov}(T(k)) &\equiv& \left<
\left( T(k)-\langle T(k)\rangle\right)
\left( T(k)-\langle T(k)\rangle \right)^\dagger
\right>  \nonumber \\
&=& s^2\sum_{l=0}^{k-1} \mathbf{G}_{k-1 k-l}\mathbf{G}_{k-1 k-l}^\dagger \ .
\label{covariance}
\end{eqnarray}

Expression (\ref{covariance}), with (\ref{Gmatrix1}) and
(\ref{iden}), gives the formal relationship between the
correlations used to construct climate networks, obtained from
the matrix $\mathrm{Cov}(T(k))$, and the transport properties
of the flow, which are contained in the flow-network matrix
$\mathbf{M}_{kk'}$ and enter into (\ref{covariance}) via
(\ref{Gmatrix1}).

\subsection*{Network construction}

From the covariance matrix we can calculate the Pearson
correlation. In terms of the matrix elements of the covariance
matrix, $\left(\mathrm{Cov}(T(k))\right)_{\vx\vy}$, the matrix
elements of the Pearson correlation matrix $\mathbf{C}(k)$ are:
\begin{equation}
\left(\mathbf{C}(k)\right)_{\vx\vy}=\frac{\left(\mathbf{Cov}(T(k))\right)_{\vx\vy}}
{\sqrt{\left(\mathbf{Cov}(T(k))\right)_{\vx\vx}\left(\mathbf{Cov}(T(k))\right)_{\vy\vy}}} \ .
\label{Pearson}
\end{equation}

As standard for climate networks, we construct correlation
networks from the symmetric and positive semi-definite matrices $\mathbf{C}(k)$. 
We threshold matrix $\mathbf{C}(k)$ to
construct a binary adjacency matrix $\mathbf{A}(k)$:
\begin{eqnarray}
\mathbf{A}(k)_{\vx\vy} &=& 1 \ \ \ \mathrm{if} \ |\mathbf{C}(k)_{\vx\vy}| \geqslant \gamma  \nonumber \\
\mathbf{A}(k)_{\vx\vy} &=& 0 \ \ \ \mathrm{if} \ |\mathbf{C}(k)_{\vx\vy}| < \gamma \ .
\label{binary}
\end{eqnarray}
Within reasonable limits the value of the threshold value
$\gamma$ below which the correlations are set to zero does not
significantly affect the result. The resulting thresholded
matrix $\mathbf{A}(k)$ is the adjacency matrix of the
correlation or climate network which is analyzed using network
measures. In the following we will tune the threshold $\gamma$
to obtain a network
with a prescribed link density. 

\subsection*{A model flow}

\begin{figure}[H]
\includegraphics[width=.8\textwidth]{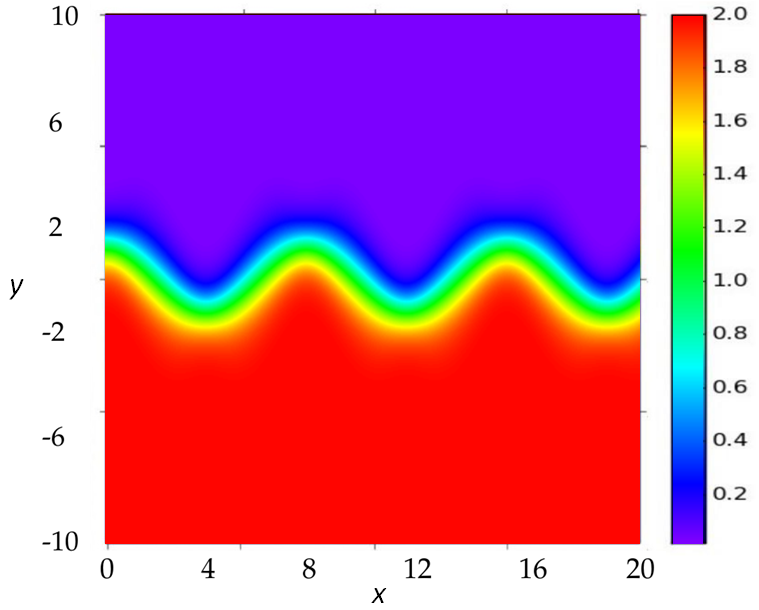}
\caption{{\bf The streamfunction for the velocity field of the meandering
flow}. It describes a jet flowing from left to right, more intense in the central meandering core.
The streamfunction is plotted here for $\nu=0$, and it is the same as for any other value of $\nu$ if $t=0$ or a
multiple of the flow period. 
Other parameters are given in the text.}
\label{meand}
\end{figure}

To illustrate the use of the formulae derived above, we choose
a \emph{meandering flow} model \cite{Samelson1991,  lopez2001} 
to construct the flow-networks.
It resembles the simplified velocity structure present in ocean
currents such as the Gulf Stream or the Kuro-Shio. 
Following
\cite{cencini1998} the streamfunction is given by:
\begin{equation}
\Psi(x,y,t) = 1- \tanh \left[\frac{y-B(t) \cos\left(m(x-ct)\right)}
{\left[1+m^2 B(t)^2\sin^2\left(m(x-ct)\right)\right]^{\frac{1}{2}}}\right],
\label{psi}
\end{equation}
where $m$ is a wave (meander) number which we set to
$2\pi/L_x$, $L_x=7.5$ and $B(t)$ is the wave amplitude, given
by $B(t) = B_0 + \nu \cos(\omega t + \theta)$. A snapshot of
the streamfunction (\ref{psi}) is plotted in Fig. \ref{meand}.
It describes a jet flowing towards the positive $x$ direction,
more intense in the central core region, and meandering in the
$y$ direction. 
A meandering flow is well-studied flow model \cite{Uleysky2007, santoleri2014}. 
Moreover, regions of the velocity field, denoted by Eq.(\ref{psi}), contain flows with more simple structure.  
Altogether this makes a meandering flow a suitable model to test a novel flow networks method. 
We fix parameters at $B_0 = 1.2$, $c = 0$,
$\omega = 0.4$, $\theta = \pi/2$, and compare results for the
static, $\nu =0$, or oscillating in amplitude, $\nu=0.7$, meander. 
In the first case particle motion in the flow is integrable
whereas in the second chaotic motions arise \cite{cencini1998,
lopez2001}. From $\Psi(x,y,t)$ the velocity field
$\vv=(v^x,v^y)$ is calculated as:
\begin{equation}
v^x(x,y,t) = -\frac{\partial \Psi(x,y,t)}{\partial y}  \ \ , \ \
v^y(x,y,t)  = \frac{\partial \Psi(x,y,t)}{\partial x}\ .
\end{equation}

\section*{Results}

In the case without advection (or advection with a constant
and homogeneous velocity field $\vv(x,y,t)=\vv_0$) Eq.
(\ref{ademodtime}) can be solved exactly and the Pearson
correlation computed. The resulting network is a fully
homogeneous graph in which every node is linked with all
neighbor nodes within a correlation length given by
$\sqrt{\kappa/b}$. In the presence of non-homogeneous advection, the network
becomes inhomogeneous with properties determined by Eq.
(\ref{covariance}) which encodes, via the propagator
$\mathbf{G}_{kk'}$, a non-trivial interplay between advection,
diffusion and decay. 
Here are some implications of our main
formula Eq. (\ref{covariance}):
\begin{itemize}
\item  In the framework of the linear ADE dynamics we are
    using here, a time-independent spatial forcing $F(\vx)$
    has no influence on the covariance matrix, as it is
    constructed from anomalies with respect to the mean. In
    the same way, white 
    noise intensity $s$ or $D$ disappears
    when normalizing the covariance to obtain the Pearson
    correlation coefficient of Eq. (\ref{Pearson}). Thus
    correlation networks become independent from the
    forcing terms present in the linear ADE Eq.
    (\ref{ademodtime}) (although these terms need to be
    present to sustain the fluctuations from which
    correlations are computed). 
    The choice 
    of the white noise in Eq.(\ref{ademodtime})
    was motivated by \cite{Hasselmann1976}, where the effect of the random weather excitation 
    on the ocean dynamics is represented by the white noise. 

\item For flow networks constructed from the transport
    matrix $\mathbf{M}_{kk'}$ (or $\mathbf{G}_{kk'}$),
    nodes are connected if there is physical transport
    between them. For networks constructed from the
    correlation (\ref{Pearson}), instead, the presence of
    the product of two propagators,  $\mathbf{G}_{k-1
    k-l}\mathbf{G}_{k-1 k-l}^\dagger$, in each term of the
    sum in Eq. (\ref{covariance}) implies that correlations
    between two nodes will be non-vanishing only if they
    receive simultaneously (at time $k$) the effect of
    fluctuations originated at the same source (at time
    $k-l$). This cannot  
    happen only by advection, because
    Lagrangian trajectories 
    are predetermined by deterministic flow model. Diffusion is needed to spread stochastic
    perturbations and let them to affect different sites.
    Thus, links between nodes in correlation networks
    constructed from transported quantities will not
    represent direct physical transport between them, but
    the susceptibility for them to be reached by
    perturbations transported (by advection and diffusion)
    from the same origin (and within a time $b^{-1}$ from
    its birth, because of the exponentially decaying
    temporal factor in  $\mathbf{G}_{kk'}$).

\item Even if for large integration time $k$ Eq. (\ref{covariance}) involves
    a large number of terms in the sum, they decrease fast
    in magnitude, and actually only the ones with $l$ such
    that $b (k-l)\Delta t < 1$ make a relevant contribution
    to the covariance or Pearson correlation at time $k$.

\item $\mathrm{Cov}(T(k))$ is a time-dependent matrix, as
    it depends on $\mathbf{G}_{kk'}$ and thus on
    $\mathbf{P}(k)$, which inherits the time-dependence on
    the velocity field $\vv(\vx,t)$. Because of the
    temporal averaging implicit in (\ref{covariance}),
    temporal scales of the velocity field faster than the
    time scale $b^{-1}$ will be averaged out from
$\mathrm{Cov}(T(k))$, but slower time-dependencies will remain and the resulting correlation
network will be a temporal network \cite{Holme2012}. 

\end{itemize}

We illustrate these general results with numerical computations
of correlations via Eqs. (\ref{covariance}) and (\ref{Pearson})
for the ADE dynamics with the meandering model flow, and
construction of the associated networks. We consider the domain
$x\in[0,20]$, $y\in[-10,10]$ with open boundary conditions and
discretize it in $N\times N=120\times 120$ nodes, so that
$\Delta x\approx 0.167$. Time step is $\Delta t=0.2$. We
nominally take the diffusion coefficient $\kappa=0.02$, but the
numerical diffusion \cite{press1988} introduced by the
discretization (\ref{dis}) is larger, $\kappa'\approx \Delta
x^2/\Delta t = 0.139$. We consider two different regimes for
the damping: $b=1$ and $b=0.05$, corresponding to lifetimes of
the perturbations much shorter ($b^{-1}=1$) than the time
scales of the flow (as given by $2\pi/\omega\approx 15.7$), or
longer ($b^{-1} = 20$). For the flow all parameters are fixed
as mentioned above, except the one giving the temporal
modulation of the meander amplitude: $\nu=0$, representing a
steady flow or  $\nu=0.7$, giving a time-dependent flow.

The network adjacency matrix $\mathbf{A}(k)$ is constructed
from Eqs. (\ref{covariance}), (\ref{Pearson}) and
(\ref{binary}). We find that using in the sum of Eq.
(\ref{covariance}) a number of terms $k=314$ for $b=1$ and
$k=942$ for $b=0.05$ (which satisfy the condition $bk\Delta
t>>1$) is sufficient to pass the spin-up period in which the
initial correlations (the first term in the right-hand-side of
Eq. \ref{correl1}) are still important, and to reach the
asymptotic statistical regime. When $\nu=0$ the flow is static,
with streamfunction plotted in Fig. \ref{meand}, and then the
network constructed from $\mathbf{A}(k)$ is also static. When
$\nu \ne 0$ the flow, and then the correlations and the
network, is periodic with period $2\pi/\omega$. For the values
used for $k$, the times $k\Delta t$ correspond to exactly 4 or
12 periods after time $t=0$ so that at these instants the
streamfunction is also the one plotted in Fig. \ref{meand}. To
highlight the spatial structures in the network we fix the
threshold $\gamma$ such that the node density is $0.075$ for
the cases with $b=0.05$, and  $0.003$ for $b=1$. Because of the
different values we 
cannot directly compare the absolute
values of the network metrics computed at different $b$. But we
will be only interested in the spatial patterns.
We have checked that, although details of the degree and clustering 
distributions vary, 
changing the link density in a factor of two does not alter 
the location of the regions of high and low values of degree and clustering with respect 
to the ones in Figs.\ref{degrees} and \ref{clustering}.

To analyze the network structure we calculate standard network
measures
\cite{dorogov2002,Boccaletti2006,Newman2003}}: 
\emph{node degree centrality}, which is the number of links
adjacent to the node, and \emph{ node clustering coefficient},
which is the fraction of triangles actually present through
that node with respect to the possible ones, given their
neighbors. The degree of the nodes in the network is plotted in
Fig. \ref{degrees} for the four combination of parameters
involving $\nu=0,0.7$ and $b=1,0.05$. Fig. \ref{clustering}
displays the corresponding clustering values.

In the static case ($\nu=0$, panels A and B of Figs.
\ref{degrees} and \ref{clustering}) the streamfunction, given
in Eq. (\ref{psi}) is constant in time, and plotted in Fig.
\ref{meand}. As expected from Eq. (\ref{covariance}) and the
discussion above, regions of high degree are not precisely
associated with strong currents. Nevertheless, when damping
rate is fast ($b=1$, Fig. \ref{degrees}A) the general spatial
structure of the degree reflects the meandering shape of the
flow. The similarity is stronger between flow and clustering
plots (Fig. \ref{clustering}A): patches of strong clustering
follow the meander structure, with high clustering usually
associated to zones of low degree, and viceversa.

\begin{figure}
\includegraphics[width=\textwidth]{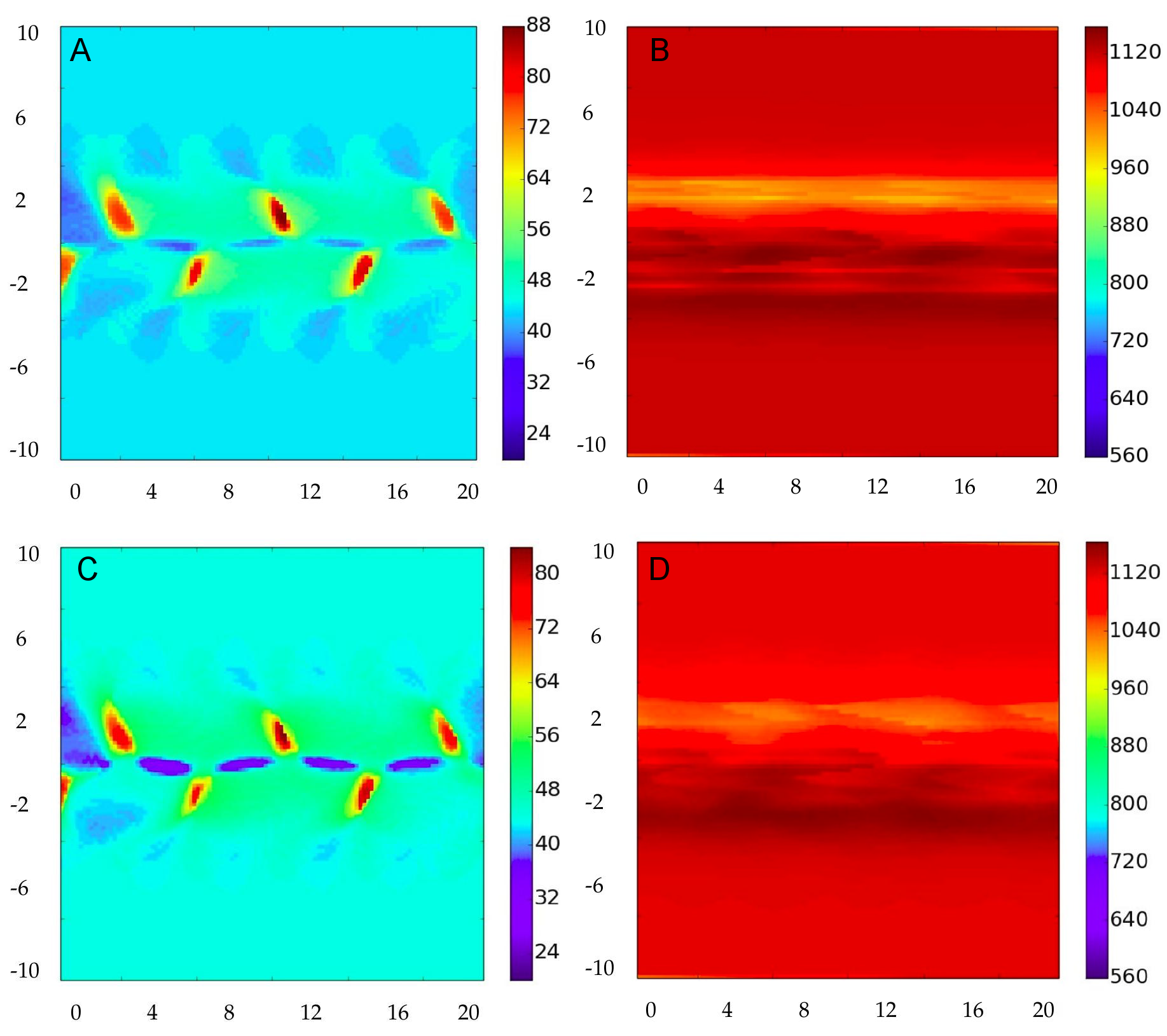}
\caption{ {\bf Node degree centrality for the correlation
networks constructed for different flows and decay rates}. The
direction $x$ is horizontal and $y$ is the vertical. Panels A
and B display the case of the static flow, $\nu=0$. C and D are
for the amplitude-changing case, $\nu=0.7$. The network for the dynamic case 
is plotted at a time after $t=0$ multiple of the
flow period. Then, for all panels the streamfunction at the
time plotted is the one shown in Fig. \ref{meand}. Panels A and
C are for the fast decay case $b=1$, and B and D are for the
slow decay, $b=0.05$, of the transported substance. Other
parameters as stated in the text.} \label{degrees}
\end{figure}

\begin{figure}
\includegraphics[width=\textwidth]{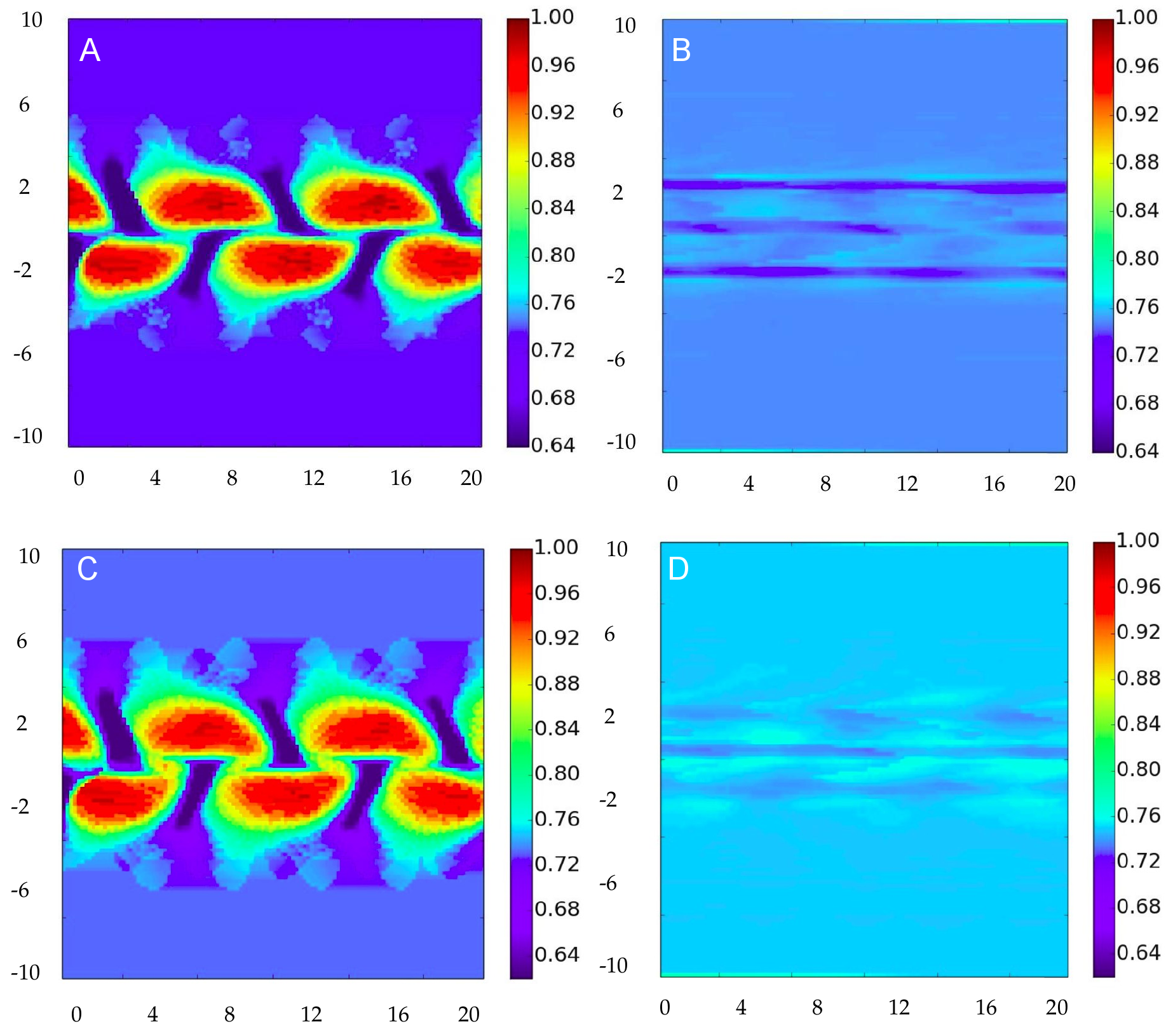}
\caption{ {\bf Node clustering coefficient for the correlation
networks constructed for different flows and decay rates}.
Panels are for the same parameters as in Fig. \ref{degrees}.}
\label{clustering}
\end{figure}

The situation completely changes for $b=0.05$ (Figures
\ref{degrees}B and \ref{clustering}B). Here both degree and
clustering become nearly homogeneous, with only some weak
structure elongated on the horizontal $x$ direction. The reason
is that now many terms corresponding to different times
contribute to the sum in Eq. (\ref{covariance}), averaging the
resulting correlations that loose spatial structure.

If we turn on now the temporal dependence of the flow,
$\nu=0.7$, little changes are seen. For the case $b=1$ (Figures
\ref{degrees}C and \ref{clustering}C) this is easy to
interpret, since as discussed above only a few terms in the sum
in Eq. (\ref{covariance}), the ones with $(k-l)b\Delta t<1$,
contribute. For them the flow stays essentially unchanged
(the time scale for changes in the flow is
$2\pi/\omega\approx 15.7 \gg b^{-1}=1$). Thus the results
should be nearly equivalent to the static case. In fact only
small increases in degree in the central parts and decreases of
degree at the maxima are seen in Fig. \ref{degrees}C with
respect to the static case Fig. \ref{degrees}A. Despite the
long-time transport properties are rather different in the
static and time-dependent case (in particular Lagrangian
transport is chaotic at $\nu=0.7$ \cite{cencini1998}) a large
damping $b$ restricts the correlations to be influenced only by
the short term dynamics, which is similar to the static case.

Making the decay rate slower ($b=0.05$, Figs. \ref{degrees}D
and \ref{clustering}D) in this dynamic case for $\nu = 0.7$ has also the
consequence of homogenizing the spatial structure, in a manner
similar to that of the static case. The structure is here
slightly more homogeneous than for $\nu=0$, because of the
additional mixing associated to the chaotic dynamics.

\section*{Discussion and outlook}

The results shown above close a gap in the theoretical
understanding of the relationship between networks constructed
from correlation functions, as usually done for climate
networks, and the underlying dynamics of the fluid transport.

A first observation is that, when the Pearson correlation is
used to establish links between nodes, correlation networks are
not sensitive to steady sources and sinks of the transported 
substances. Also the normalization in Eq. (\ref{Pearson})
eliminates the dependence on fluctuation intensity. As a
consequence in geophysical contexts, one cannot 
look into
climate networks for information about these processes. Note
that this implication is strictly valid only for the linear ADE
dynamics in Eq. (\ref{ademodtime}) and will not apply to
dynamics involving nonlinear processes (plankton dynamics,
vorticity, ...). Also, it may not hold when nonlinear measures
of statistical dependence, such as mutual information,
information transfer \cite{Deza2015chaos,Deza2015} or event
synchronization \cite{Malik2011} replace the correlation
function.

Another important point, evident from Eq. (\ref{covariance}),
is that the relationship between the correlation network,
constructed from $\mathbf{C}(k)$ and the underlying flow
transport network (characterized by $\mathbf{M}_{kk'}$ or
$\mathbf{G}_{kk'}$) is not direct, since the correlation
expression involves a sum over time, and each term involves the
product of two propagators, meaning that correlated nodes are
not the ones connected by the flow, but the ones affected
within a time $b^{-1}$ by perturbations coming from a common
origin. 
It is straightforward to repeat the calculations for the case in which a colored noise correlation is used for $\epsilon(k)$.  
The result is that correlated nodes are the ones affected 
by perturbations coming from locations within the same correlation length and time of the noise. 
In consequence, patterns of degree or of other network
measures are related to flow patterns in a rather indirect way, 
as Figs. \ref{degrees} and \ref{clustering} confirm. Note
that this result relies strongly on considering the {\sl
equal-time correlation}. In cases in which a {\sl time-lagged
correlation} is used
\cite{Yamasaki2008,Molkenthin2014a,Zhou2015}, the resulting
network would be more associated to fluid transport occurring
between nodes during the selected temporal lag. Also, our
analysis in this paper is restricted to the ADE dynamics
implemented by Eq. (\ref{ademodtime}), which considers only
material transport. Our conclusions may not apply to climate
networks constructed from variables involving wave propagation
(Kelvin, Rossby, ...), such as sea surface height or
geopotential \cite{Arizmendi2014}.

From the numerical results presented here it is seen that one
of the parameters having the largest impact on the network
topology, in fact more than the flow geometry or temporal
variability, is the characteristic time scale of perturbation
damping (here represented by the decay rate $b$). This
important parameter would then have to be taken into account
when investigating the structure of climate networks constructed
from observed or analyzed data.

In summary we have elucidated, in the context of ADE dynamics,
general relationships between correlation and flow networks,
overcoming some restrictions of previous approaches,
\cite{Molkenthin2014a}. 
Moreover, flow networks are further applicable, 
for instance, to study changes in flow behavior 
\cite{Gao201674, Garaboa-Paz2015}.  
All in all, the methods above can, in
principle, be applied in other contexts, in which temporal networks
\cite{Holme2012,Masuda2013,Tupikina2014} 
are used in order to study transport process, so the present
framework can be useful to investigate
different complex systems.

\section*{Acknowledgments}
We would like to thank Henk
Dijkstra, Frank Hellman, Alexis Tantet for helpful and interesting discussions.
Also we would like to acknowledge EC-funding through the Marie-Curie ITN LINC project (P7-PEOPLE-2011-ITN, grant No.289447), 
and FEDER and MINECO (Spain) through project ESCOLA (TM2012-39025-C02-01).

\bibliography{flow_networks_ifisc_pik}

\end{document}